\documentclass[final,nospthms]{svjour3_like}

\usepackage{amssymb}
\usepackage{graphicx}

\begin{document}

\title{Generalized modeling of ecological population dynamics}

\author{Justin D.~Yeakel         \and
        Dirk Stiefs \and
        Mark Novak \and
        Thilo Gross
}

\institute{
Justin D.~Yeakel \at
Department of Ecology and Evolutionary Biology\\
1156 High Street \\
University of California \\
Santa Cruz, CA 95064 \\
USA\\
\email{jdyeakel@gmail.com} \\         
\and
Dirk Stiefs \at
Max-Planck Institute for the Physics of Complex Systems
N\"othnitzer Str.~38 \\
01187 Dresden \\
Germany \\
\email{stiefs@pks.mpg.de} \\ 
\and
Mark Novak \at
Long Marine Lab \\
100 Shaffer Road \\
Santa Cruz, CA 95060 \\
USA\\
\email{mnovak1@ucsc.edu} \\ 
\and
Thilo Gross \at
Max-Planck Institute for the Physics of Complex Systems
N\"othnitzer Str.~38 \\
01187 Dresden \\
Germany \\
\email{thilo.gross@physics.org} \\ 
}

\date{\today}

\maketitle
\begin{abstract}
Over the past years several authors have used the approach of generalized modeling
to study the dynamics of food chains and food webs. Generalized models come close 
to the efficiency of random matrix models, while being as directly interpretable 
as conventional differential-equation-based models.  
Here we present a pedagogical introduction to the approach of generalized modeling. 
This introduction places more emphasis on the underlying concepts of generalized 
modeling than previous publications. 
Moreover, we propose a shortcut that can significantly accelerate the formulation of 
generalized models and introduce an iterative procedure that can be used to refine existing generalized models by integrating 
new biological insights.  
 
\end{abstract}

\section{Introduction}
\label{sec:1}
Ecological systems are fascinating because of their complexity. 
Not only do ecological communities harbor a multitude of different species, but even the interaction of just two individuals can be amazingly complex. 
For understanding ecological dynamics this complexity poses a considerable challenge. 
In conventional mathematical models, the dynamics of a system of interacting species are described by a specific set of ordinary differential equations (ODEs).
Because these equations are formulated on the level of the population, all complexities arising in the interaction of individuals must be cast into specific functional forms. 
Indeed, several important works in theoretical ecology present derivations of functional forms that include certain types of individual-level effects \cite{holling,Rosenzweig1971,Berryman1981,Getz:1984p2844,Fryxell:2007p1462}. 
Although these allow for a much more realistic representation than, say, simple mass-action models, they cannot come close to capturing all the complexities existing in the real system.
Even if detailed knowledge of the interactions among individuals were available and could be turned into mathematical expressions, these would arguably be too complex to be conducive to a mathematical analysis. 
In this light the functional forms that are commonly used in models can be seen as a compromise, reflecting the aim of biological realism, the need to keep equations simple, and often the lack of detailed information.    

Because of the many unknowns that exist in ecology, it is desirable to obtain results that are independent of the specific functional forms used in the model. 
This has been achieved by a number of studies that employed general models, in which at least some functional forms were not specified
\cite{Gardner1970,May1972,DeAngelis1975,Murdoch1975,Levin1977,Murdoch1977,Wollkind1982}.
These works considered not specific models, but rather classes of models comprising simple, commonly used, functions, as well as the whole range of more complex alternatives.

That ecological systems can be analyzed without restricting the interactions between populations to specific functional forms is in itself not surprising--in every mathematical analysis the objects that are analyzed can be treated as unknown. 
The results of the analysis will then depend on certain properties of the unknown objects. 
In a general ecological model we thus obtain results that link dynamical properties of the model, e.g.~the presence of predator-prey oscillations to properties of the (unknown) functions describing certain processes, e.g. the slope of the functional response evaluated at a certain point.
Accordingly, the analysis of general models reveals the decisive properties of the functional forms that have a distinctive impact on the dynamics.
Whether such results are ecologically meaningful depends crucially on our ability to attach an ecological interpretation to the decisive properties that are identified.

In the present paper we specifically consider the approach called \emph{generalized modeling}.
This approach constitutes a procedure by which the local dynamics in models can be analyzed in such a way that the results are almost always interpretable in the context of the application. 
Generalized modeling was originally developed for studying food chains \cite{Gross:2004p2428,GrossTheorBiol,Gross:2005p2432} and was only later proposed as a general approach to nonlinear dynamical systems \cite{Gross:2006p2337}.
Subsequently, generalized modeling was used in systems biology, where it is sometimes called structural-kinetic modeling \cite{RalfSteuer08082006,Steuer2007,Zumsande2010a,reznik_stability_2010} and is covered in recent reviews \cite{Steuer2007b,Sweetlove2008,Jamshidi2008,Steuer2009,Rodriguez2009,Schallau2010}. 
In ecology generalized models have been employed in several recent studies \cite{Stiefs2010,Gross:2009p2158,Equivalence,George,Baurmann}, for instance for explorating the effects of food-quality on producer-grazer systems \cite{Stiefs2010} and for identifying stabilizing factors in large food webs \cite{Gross:2009p2158}. 
The latter work demonstrated that the approach of generalized modeling can be applied to large systems comprising 50 different species and billions of food web topologies. 
   
In the present paper we present a pedagogical introduction to generalized modeling and explain the underlying idea on a deeper level than previous publications.
Furthermore, we propose some new techniques that considerably facilitate the formulation and analysis of generalized models.
The approach is explained using a series of ecological examples of increasing complexity, including a simple model of omnivory that has so far not been analyzed by generalized modeling.

We start out in Sec.~\ref{secMath} with a brief introduction to fundamental concepts of dynamical systems theory. In Sec.~\ref{secGrowth}, we introduce generalized modeling by considering the example of a single population. 
In contrast to previous generalized analyses of this system we use a shortcut that accelerates the formulation of generalized models.
This shortcut is also used in Sec.~\ref{secPred}, where we apply generalized modeling to a predator-prey system. 
Our final example, shown in Sec.~\ref{secOmni}, is a simple omnivory scenario involving three species. 
This example already contains all of the difficulties that are also encountered in larger food webs. 

\section{Local analysis of dynamical systems}
\label{secMath}
Generalized modeling builds on the tools of nonlinear dynamics and dynamical systems theory. 
Specifically, information is typically extracted from generalized models by a local bifurcation analysis. 
Mathematically speaking, a bifurcation is a qualitative transition in the long-term dynamics of the system, such as the transition from stationary (equilibrium) to oscillatory (cyclic) long-term dynamics. The corresponding critical parameter value at which the transition occurs is called the \emph{bifurcation point}. 
In this section we review the basic procedure for locating bifurcation points in systems of coupled ODEs. 
This analysis is central to the exploration of both generalized and conventional models and is also covered in many excellent text books, for instance \cite{kuznetsov2004,Guckenheimer2002}.   

In the following we consider systems of $N$ coupled equations 
\begin{equation} 
\frac{\rm d}{\rm dt} x_i = f_i(\mathbf{x})
\label{eqODE}
\end{equation}
where $\mathbf{x}=(x_1,\ldots,x_N)$ is a vector of variables and $f(\mathbf{x})$ is a vector-valued function. 
In population dynamics, each $x_i$ typically corresponds to a population, representing the abundance, biomass, or biomass density. 

The simplest form of long-term behavior that can be observed in systems of ODEs is stationarity. 
In a \emph{steady state} $\mathbf{x}^*$ the right hand side of the equations of motion vanishes,
\begin{equation}
\frac{\rm d}{\rm dt} {x_i}^*=0
\end{equation}  
for all $i$. Therefore, a system that is placed in a steady state will remain at rest for all time.  

Stationarity alone does not imply that a state is a stable equilibrium.
A system that is perturbed slightly from the steady state may either return to the steady state asymptotically in time or depart from the steady state entirely.
For deciding whether a steady state is stable against small perturbations, we consider the local linearization of the system around the   
steady state, which is given by the corresponding Jacobian $\rm \bf J$, an $N\times N$ matrix with 
\begin{equation}
J_{ij} = \left.\frac{\partial }{\partial x_j} f_i(\mathbf{x})\right|_*
\end{equation}
where $|_*$ indicates that the derivative is evaluated in the steady state.
 
Because the Jacobian is a real matrix, its eigenvalues are either real or form complex conjugate eigenvalue pairs. 
A given steady state is stable if all eigenvalues of the corresponding Jacobian $\rm \bf J$ have negative real parts.
When the function $f(\mathbf{x})$ is changed continuously, for instance by a gradual change of parameters on which $f(\mathbf{x})$ depends, 
the eigenvalues of the corresponding Jacobian change continuously as well. 

Local bifurcations occur when a change in parameters causes one or more eigenvalues to cross the imaginary axis of the complex plane. 
In general, this happens in either of two scenarios: In the first scenario, a real eigenvalue crosses the imaginary axis, causing a \emph{saddle-node bifurcation}. In this bifurcation two steady states collide and annihilate each other. 
If the system was residing in one of the steady states before the transition, the variables typically change rapidly while the system approaches some other attractor. 
In ecology crossing a saddle-node bifurcation backwards can, for instance, mark the onset of an Allee effect. 
In this case one of the two steady states emerging from the bifurcation is a stable equilibrium, whereas the other is an unstable saddle, which marks the tipping point between long-term persistence and extinction.

In the second scenario, a complex conjugate pair of eigenvalues crosses the imaginary axis, causing a \emph{Hopf bifurcation}. 
In this bifurcation the steady state becomes unstable and either a stable limit cycle emerges (supercritical Hopf) or an unstable limit cycle 
vanishes (subcritical Hopf). The supercritical Hopf bifurcation marks a smooth transition from stationary to oscillatory dynamics. 
A famous example of this bifurcation in biology is found in the Rosenzweig-MacArthur model \cite{Rosenzweig1963}, where enrichment leads to destabilization of a steady state in a supercritical Hopf bifurcation.
By contrast, the subcritical Hopf bifurcation is a catastrophic bifurcation after which the system departs rapidly from the neighborhood of the steady state.

In addition to the generic local bifurcation scenarios, discussed above, degenerate bifurcations can be observed if certain symmetries exist in the system. 
In many ecological models one such symmetry is related to the unconditional existence of a steady state at zero population densities. 
If a change of parameters causes another steady state to meet this extinct state, then the system generally undergoes a \emph{transcritical bifurcation} in which the steady states cross and exchange their stability. 
The transcritical bifurcation is a degenerate form of the saddle-node bifurcation and is, like the saddle-node bifurcation, characterized by the existence of a zero eigenvalue of the Jacobian.     
 
\section{Density Dependent Growth of a Single Species}
\label{secGrowth}
In the following we demonstrate how the approach of generalized modeling can be used to find local bifurcations in general ecological models.
We start with the simplest example: the growth of a single population X. 
A generalized model describing this type of system can be written as
\begin{equation}
\frac{\rm d}{\rm dt}X = S(X) - D(X)
\label{eqSinglePop}
\end{equation}
where $X$ denotes the biomass or abundance of population X, $S(X)$ models the intrinsic gain by reproduction, and $D(X)$ describes the loss due to mortality. 
In the following we do not restrict the functions $S(X)$ and $D(X)$ to specific functional forms. 

In the following, we consider all positive steady states in the whole class of systems described by Eq.~(\ref{eqSinglePop}) and ask which of those states are stable equilibria. 
For this purpose we denote an arbitrary steady state of the system as $X^*$. 
We emphasize that $X^*$ is not a placeholder for any specific steady state that will later be replaced by numerical values,
but should rather be considered a formal surrogate for every single steady state that exists in the class of systems. 

For finding the decisive factors governing the stability of $X^*$ we compute the Jacobian 
\begin{equation}
\mathbf{J^*} = \left.\frac{\partial S}{\partial X} \right|_{*} - \left.\frac{\partial D}{\partial X} \right|_{*}.
\end{equation}
Because evaluated in the steady state, the two terms appearing on the right hand side of this equation are no longer functions but 
constant quantities. 
We could therefore formally consider these terms as unknown parameters. 
While mathematically sound, parameterizing the Jacobian in this way leads to parameters that are hard to interpret in the context of the model and are therefore not conducive to an ecological analysis. 
We therefore take a slightly different approach and use the identity   
\begin{equation}
\label{eqIdentity}
\left.\frac{\partial F}{\partial X} \right|_{*} =\frac{F^*}{X^*} \left.\frac{\partial \log F}{\partial \log X}\right|_{*},
\end{equation}
where $F$ is an arbitrary positive function and we abbreviated $F(X^*)$ by $F^*$. The identity, Eq.~(\ref{eqIdentity}), holds for all $F^*>0$ and $X^*>0$; its derivation is shown in App.~A. 
Substituting the identity into the Jacobian, we obtain
\begin{equation}
\label{eqPrelimJac}
\mathbf{J^*} =\frac{S^*}{X^*} s_{\rm x} - \frac{D^*}{X^*} d_{\rm x}.
\end{equation} 
where
\begin{equation}
\label{eqDefSx}
s_{\rm x} := \left.\frac{\partial \log S}{\partial \log X}\right|_{*},
\end{equation}
\begin{equation}
\label{eqDefDx} 
d_{\rm x} :=\left.\frac{\partial \log D}{\partial \log X}\right|_{*}.
\end{equation}

We note that $S^*/X^*$ and $D^*/X^*$ denote \emph{per-capita} gain and loss rates, respectively.
Because the gain and loss have to balance in the steady state we can define 
\begin{equation}
\label{eqDefAlpha}
\alpha := \frac{S^*}{X^*} = \frac{D^*}{X^*}.
\end{equation}
The parameter $\alpha$ can be interpreted as a characteristic timescale of the population dynamics. 
If $X$ measures abundance then this timescale is the per-capita mortality rate or equivalently the per-capita birth rate, 
or in other words, the inverse of an individual's life expectancy.
If $X$ is defined as a biomass then $\alpha$ denotes the biomass turnover rate. 
Using $\alpha$ the Jacobian can be written as
\begin{equation}
\mathbf{J^*} =\alpha \left( s_{\rm x} - d_{\rm x} \right).
\end{equation}

\begin{figure}
   \centering
   \includegraphics[width=0.75\textwidth]{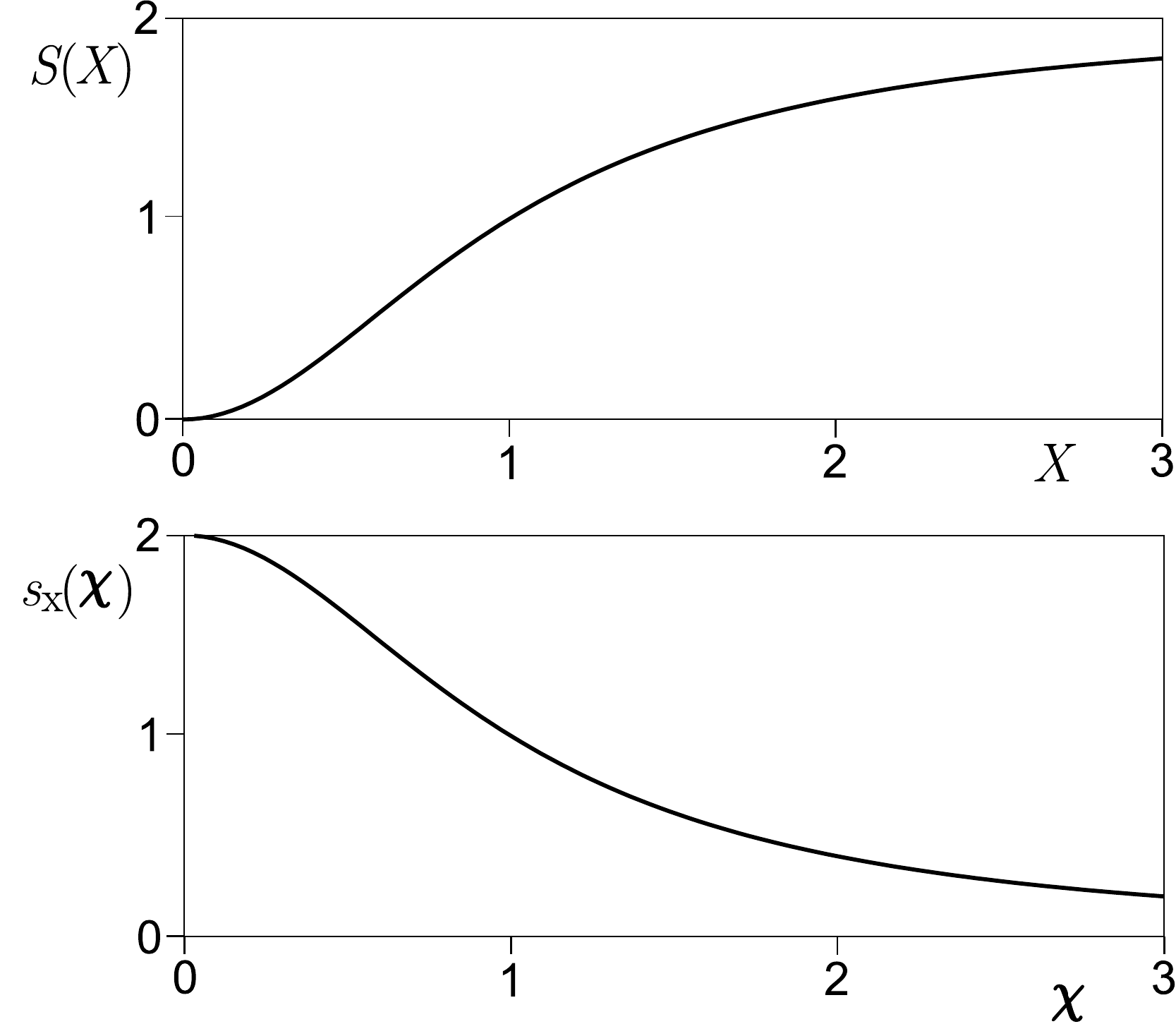}
      \caption{{\bf A}. In the specifc example system  a reporoduction rate, $S(X)$ of the form of a Holling type-III functional response, $aX^2/(k^2+X^2)$, is assumed. This function starts out quadratically at low values of the population density $X$, but saturates as $X$ increases.  
      {\bf B}. The corresponding elasticity, $s_{\rm x}$, is close to two near the quadratic regime $\chi=k/X^*\approx 0$, but approaches zero as saturation sets in.}
      \label{fig:1}
\end{figure}

Let us now discuss the interpretation of the other two parameters $s_{\rm x}$ and $d_{\rm x}$. For this purpose, note that these parameters are defined as logarithmic derivatives of the original functions. 
Such parameters are also called \emph{elasticities}, because they provide a nonlinear measure for the sensitivity of the function to variations in the argument. 
For any power-law $aX^p$ the corresponding elasticity is $p$. For instance all constant functions have an elasticity of 0, all linear functions an elasticity of 1, and all quadratic functions have an elasticity of 2. This also extends to decreasing functions such as $a/X$ for which the corresponding elasticity is -1.
For more complex functions the value of the elasticity can depend on the location of the steady state.
However, even in this case the interpretation of the elasticity is intuitive. For instance the Holling Type-II functional response is linear for low prey density and saturates for high prey density \cite{holling}. The corresponding elasticity is approximately 1 in the linear regime, but asymptotically decreases to 0 as the predation rate approaches saturation. A similar comparison for the Holling type-III function is shown in Fig.~\ref{fig:1}.

Elasticities are used in several scientific disciplines because they are directly interpretable and can be easily estimated from data \cite{Fell1985}.
In particular we emphasize that elasticities are defined in the state that is observed in the system under consideration, and thus 
do not require reference to unnatural situations, such as half-maximum values or rates at saturation that often cannot be observed directly. 
We note that in previous publications the elasticities have sometimes been called exponent parameters and have been obtained by a normalization procedure. 
In comparison to this previous procedure the application of Eq.~(\ref{eqIdentity}), proposed here, provides a significant shortcut. 

We now return to the discussion of the example system. 
So far we have managed to express the Jacobian determining the stability of all steady states by the three parameters $\alpha$, $s_{\rm x}$, and $d_{\rm x}$.
Because this simple example contains only one variable, the Jacobian is a 1-by-1 matrix. 
Therefore, the Jacobian has only one eigenvalue which is directly 
\begin{equation}
\lambda=\alpha \left( s_{\rm x} - d_{\rm x} \right).
\end{equation}
The steady state under consideration is stable if $\lambda<0$, or equivalently
\begin{equation}
\label{eqstabcond}
s_{\rm x}<d_{\rm x}.
\end{equation}
In words: In every system of the form of Eq.~(\ref{eqSinglePop}) a given steady state is stable whenever the elasticity of the mortality in the steady state exceeds the elasticity of reproduction. 

A change in stability occurs when the elasticities of gain and loss become equal, 
\begin{equation}
\label{bifcond}
s_{\rm x}=d_{\rm x}.
\end{equation}
If this occurs the eigenvalue of the Jacobian vanishes and the system undergoes a saddle-node bifurcation. 

For gaining a deeper understanding of how the generalized analysis relates to conventional models it is useful to consider a specific example. 
One model that immediately comes to mind is logistic growth, which is characterized by linear reproduction and quadratic mortality. 
However, based on our discussion above, it is immediately apparent that linear reproduction must correspond to $s_{\rm x}=1$ and quadratic mortality to $d_{\rm x}=2$. 
Without further analysis we can therefore say that steady states found for a single population under logistic growth must always be stable regardless of the other parameters.  

\begin{figure*}
   \centering
   \includegraphics[width=1\textwidth]{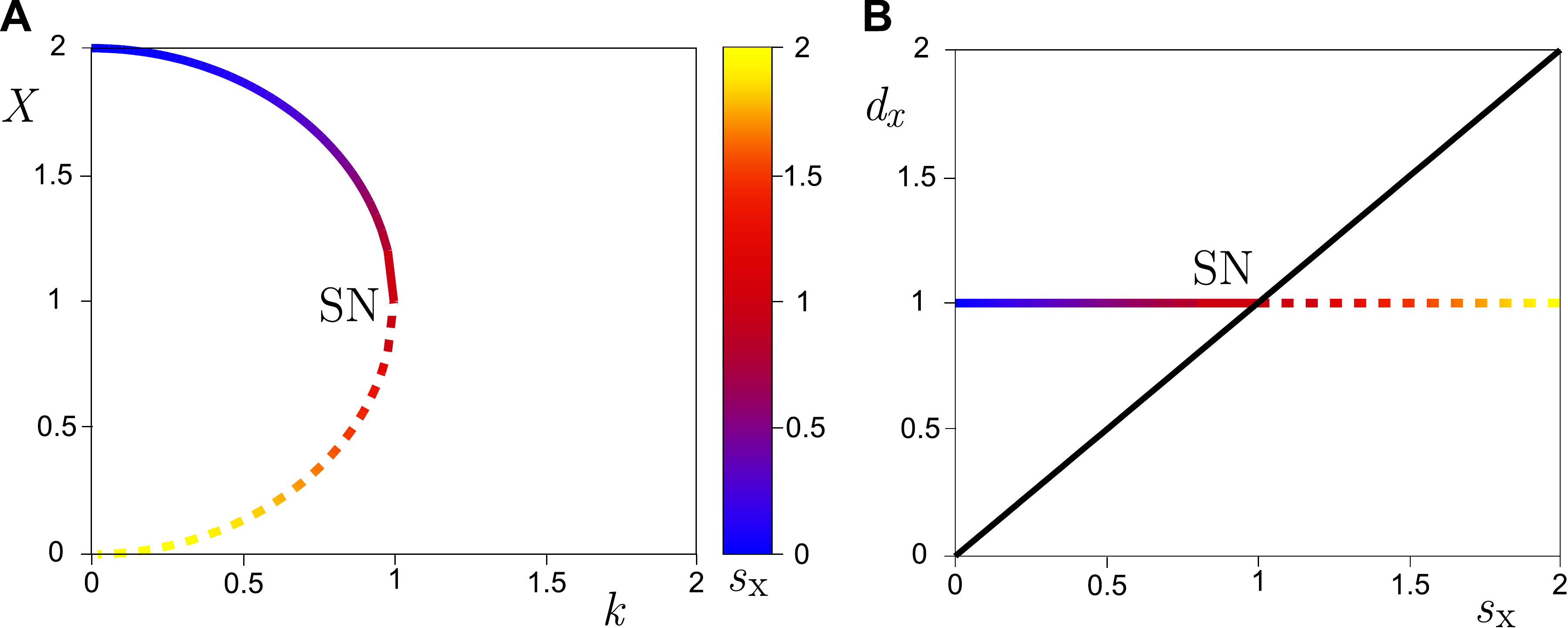}
      \caption{{\bf A}. Bifurcation diagram of a specific example (Eq.~\ref{eqsingle}). The lines correspond to the locations of steady states, which are stable equilibria (solid) or saddles (dashed). The color encodes the elasticity of growth, $s_{\rm x}$, in the respective steady states. The figure confirms the our expectation from the generalized model that steady states are stable whenever $s_{\rm x}<d_{\rm x}$, where $d_{\rm x}=1$ in the specific example. The two steady states vanish in a saddle node bifurcation, which occurs at $s_{\rm x}=d_{\rm x}$. 
{\bf B}. The correspondence between generalized and specific model can be seen explicitly by mapping the steady states from the specific model into the generalized parameter plane. In this plane the stable and unstable states are separated by the stability boundary (black line, Eq.~\ref{bifcond}) at which the saddle-node bifurcation occurs}
\label{fig:2}
\end{figure*}

A more interesting example is obtained when one assumes a reproduction rate following a Holling type-III kinetic and linear mortality,
\begin{equation}
\label{eqsingle}
\frac{\rm d}{\rm dt}X = \frac{aX^2}{k^2+X^2} - bX,
\end{equation}
where $a$ is the growth rate at saturation, $k$ is the half-saturation value of growth, and $b$ is the mortality rate. This example system 
can be investigated by explicit computation of steady states and subsequent stability and bifurcation analysis. 
This procedure is shown in most textbooks on mathematical ecology and is hence omitted here. 
For the present example the conventional analysis reveals that, for high $k$ only a trivial equilibirum at zero population density exists, so that the population becomes extinct deterministically (Fig.~\ref{fig:2}A). As $k$ is reduced, a saddle-node bifurcation occurs, which marks the onset of an Allee effect. 
In the bifurcation a stable non-trivial equilibrium and an unstable saddle point are created. Beyond the bifurcation a population can persist if its initial abundance is above the saddle point. In this case the population asymptotically approaches the stable equilibrium. By contrast, a population which is initially below the saddle point declines further and approaches the trivial (extinct) equilibrium. 

For comparing the results from the specific analysis to the generalized model, we compute the elasticities that characterize the steady states found in the specific model. Because the mortality rate is assumed to be linear, we know $d_{\rm x} = 1$.
The elasticity of the growth function can be found by applying Eq~(\ref{eqIdentity}) to the known growth function of the specific model.
This yields
\begin{equation}
\label{eqsx}
s_{\rm x} = \frac{2}{1 + \chi^2}
\end{equation}
where $\chi = X^*/k$. A detailed derivation of this relationship using a normalization procedure instead of the shortcut Eq.~(\ref{eqIdentity}), is given in \cite{GrossTheorBiol}. 
Equation (\ref{eqsx}) shows that the elasticity of growth is $s_{\rm x}\approx 2$ for $X^*\ll k$, but approaches $s_{\rm x}=0$ in the limit $X^*\gg k$ (Fig.~\ref{fig:1}).

In Fig.~\ref{fig:2}A we have color-coded the growth elasticity of steady states visited by the system as $k$ is changed. 
We note that the saddle-node bifurcation occurs at $s_{\rm x}=d_{\rm x}=1$, conforming to our expectation from the generalized model. 
Moreover, in the unstable saddles we find $s_{\rm x}>d_{\rm x}$, whereas the stable equilibria are characterized by $s_{\rm x}<d_{\rm x}$, which is in agreement with Eq.~(\ref{eqstabcond}). 

We can now map the steady states found in the specific model into the generalized parameter plane spanned by the elasticities $s_{\rm x}$ and $d_{\rm x}$ (Fig.~\ref{fig:2}B). 
Because $d_{\rm x}=1$ in the specific example, irrespective of $X^*$, all steady states end up on a single line in the generalized diagram. 
Other areas of the bifurcation diagram, not visited by the specific example, correspond to other models that assume other functional forms for the mortality. 
In this diagram the folding back of the branch of steady states, which characterizes the saddle-node bifurcation in the specific model, is not visible. 
However, from the bifurcation condition, Eq.~(\ref{bifcond}), we know that this bifurcation must occur as the diagonal line in the diagram is crossed. 

The comparison of the two bifurcation diagrams in Fig.~\ref{fig:2} highlights the differences between generalized and conventional modeling. 
In the conventional model different numbers of steady states are found depending on the specific values of parameters that are assumed.
Moreover, for a given set of parameter values multiple steady states can coexist that differ in their stability properties. 
Because the generalized model 
comprises a whole class of specific models a single set of generalized parameters corresponds to an infinite number of different 
steady states, found in different specific models. However, the solution branches of this family of models have been unfolded such 
that all steady states corresponding to the same set of generalized parameters must have the same stability properties. 

It is apparent that for a given specific example the conventional analysis reveals more detailed insights than the generalized analysis. For instance the presence of the Allee effect that is directly evident in the conventional bifurcation diagram, Fig.~\ref{fig:2}A, can only be inferred indirectly from 
the presence of the saddle-node bifurcation in the generalized analysis, Fig.~\ref{fig:2}B.  
However, the conventional analysis provides insights only into the dynamics of the specific example, whereas the generalized analysis reveals results that are valid for a whole class of models and are hence robust against uncertainties in the specific model.

A major advantage of the generalized model is that results are obtained without explicit computation 
of steady states. In the conventional model that we discussed in this section, steady states can be computed analytically.
However, even for slightly more complex models this computation becomes infeasible as it involves (under the best circumstances) factorization 
of large polynomials. Also the numerical computation of steady states poses a serious challenge for which no algorithm with guaranteed 
convergence is known. Because generalized modeling avoids the explicit computation of steady states, the approach can be scaled to much larger 
networks. The additional complications that arise in the generalized modeling of larger systems and their resolution are the subject of the subsequent sections.

\section{Predator-Prey Dynamics}
\label{secPred}
In our second example, we consider a slightly more complex system where intra- and inter-specific interactions are considered.
Departing from the single species model, we introduce a predator Y whose growth is entirely dependent on ${\rm X}$. This leads to the generalized model
\begin{eqnarray}
&& \frac{{\rm d}}{{\rm dt}}X = S(X) - D(X) - F(X,Y), \label{eq:sysprey}\\\nonumber\\
&& \frac{{\rm d}}{{\rm dt}}Y = \gamma F(X,Y) - M(Y), \label{eq:syspred}
\end{eqnarray}
where $S(X)$ and $D(X)$ describe the reproduction and mortality of the prey X, the function $F(X,Y)$ models the interaction of X with the predator Y, $M(Y)$ is the mortality of Y, and $\gamma$ is a constant conversion efficiency. 
By linearizing around an unknown steady state $(X^*,Y^*)$ and using the identity Eq.~(\ref{eqIdentity}) we obtain the Jacobian 
\begin{equation} 
\label{eqPrelimJac2}
\mathbf{J^*} =
\left(
\begin{array}{ccc}
\frac{S^*}{X^*} s_{\rm x} - \frac{D^*}{X^*} d_{\rm x} - \frac{F}{X^*} f_{\rm x} && - \frac{F^*}{X^*} f_{\rm y} \\ \\
\gamma \frac{F^*}{Y^*} f_{\rm x} && \gamma \frac{F^*}{Y^*}f_{\rm y} - \frac{M^*}{Y^*} m_{\rm y}
\end{array}
\right),
\end{equation}
where
\begin{eqnarray}
&& s_{\rm x} := \left.\frac{\partial \log S}{\partial \log X}\right|_*, ~~ d_{\rm x} := \left.\frac{\partial \log D}{\partial \log X}\right|_*, ~~ f_{\rm x} :=  \left.\frac{\partial \log F}{\partial \log X}\right|_*, \nonumber\\
&& f_{\rm y} := \left. \frac{\partial \log F}{\partial \log Y}\right|_*, ~~ m_{\rm y} := \left. \frac{\partial \log M}{\partial \log Y}\right|_*. 
\label{eq:predpreydefs}
\end{eqnarray}
As a next step we absorb the steady state abundances $X^*$, $Y^*$, rates $S^*$, $D^*$, $F^*$, $M^*$, and the constant $\gamma$ into a set of scale 
parameters. 
In doing so we have to take care to satisfy the demands of stationarity for every variable. Let us first consider the predator. The corresponding equation of motion, 
Eq.~(\ref{eq:syspred}), implies
\begin{equation}
\label{predcosistency}
\gamma F^* - M^* =0.
\end{equation}
Analogous to the example from the previous section, we can therefore define  
\begin{equation}
\alpha_y := \gamma \frac{F^*}{Y^*} = \frac{M^*}{Y^*}.
\end{equation}
which automatically satisfies Eq.~(\ref{predcosistency}) but does not restrict the per-capita rates otherwise.
As in the previous example, the $\alpha_y$ can be interpreted as a characteristic time scale, now describing the predator population. 
 
An additional complication is encountered for the prey because the stationarity condition corresponding to Eq.~(\ref{eq:sysprey}) contains three terms, 
\begin{equation}
S^*-D^*-F^* = 0.
\end{equation}
In such a case it is almost always advantageous to first define a characteristic time scale $\alpha$, which equals the sum of all per-capita gains and the sum of all per-capita losses. For instance in the present system we define 
\begin{equation}
\alpha_x := \frac{S^*}{X^*} = \frac{D^*}{X^*} + \frac{F^*}{X^*},
\end{equation} 
where all gains appear on the left side of the equals sign and all losses appear on the right. Defining the turnover in this way, guarantees stationarity 
of the state under consideration. 
However, because the terms $D^*/X^*$ and $F^*/X^*$ appear independently in the Jacobian, defining $\alpha_x$ is not sufficient for replacing all occurences of the per-capita rates in the Jacobian. We therefore define a second parameter 
\begin{equation}
\beta := \frac{1}{\alpha_x} \frac{D^*}{X^*},
\end{equation}
and its complement
\begin{equation}
\bar{\beta} := 1 - \beta = \frac{1}{\alpha_x} \frac{F^*}{X^*}, \label{eq:xdef}
\end{equation}
which describe the branching of the biomass flow. In words, $\beta$ denotes the proportion of the total loss of the prey that occurs due to mortality, whereas $\bar{\beta}$ denotes the proportion of the total loss of the prey that occurs due to predation.

In general, the same strategy for defining branching parameters can be applied to equations containing any number of terms.
For each variable, first define a parameter $\alpha$, which denotes the total turnover rate, separating gain and loss terms and identifying the characteristic timescale of a species.
Branching parameters are then assigned to any number of terms that define the relative contribution of the individual gains and losses to the total turnover within a system. We emphasize that by design the branching parameters arising from each differential equation add up to one, which is necessary for consistency. 
 
Returning to the generalized predator-prey system, we substitute the scale and branching parameters into Eq.~(\ref{eqPrelimJac2}), which yields the Jacobian
\begin{equation}
\label{eqJac2} 
\mathbf{J^*} =
\left(
\begin{array}{ccc}
\alpha_{\rm x}\left(s_{\rm x} - \beta  d_{\rm x} - \bar{\beta} f_{\rm x} \right)  && -\alpha_{\rm x} \bar{\beta} f_{\rm y} \\
\alpha_{\rm y} f_{\rm x}  && \alpha_{\rm y} \left( f_{\rm y} -  m_{\rm y}\right)
\end{array}
\right).
\end{equation}

In contrast to the system from the previous section, the Jacobian is now a 2-by-2 matrix. For this Jacobian the eigenvalues can 
still be computed analytically.   
However, analytical eigenvalue computation is tedious already for systems with 3 variables, and in general impossible for systems with more than 4 variables.
Nevertheless, analytical results can be obtained even for larger systems by deriving testfunctions that directly test for bifurcations, without an intermediate computation of eigenvalues.

Saddle-node bifurcations occur when a single real eigenvalue crosses the imaginary axis. Therefore, a zero eigenvalue must be present in a saddle-node bifurcation. This implies that the product of all eigenvalues must vanish in this bifurcation. Because the product of all eigenvalues equals the determinant of a matrix, we can locate saddle-node bifurcations by demanding that the determinant of the Jacobian, ${\rm det~{\bf J^*}}>0$, vanishes. 
For the present example this yields the condition
\begin{equation}
s_{\rm x} = \frac{(\beta d_{\rm x}+\bar{\beta} f_{\rm x})(f_{\rm y}-m_{\rm y})-\bar{\beta}f_{\rm x}f_{\rm y}}{f_{\rm y}-m_{\rm y}}.
\end{equation}

For finding the Hopf bifurcations we note that the trace of a matrix (the sum of diagonal elements) is identical to the sum of the eigenvalues. 
For a two-dimensional system this implies that the trace of the Jacobian, ${\rm tr~{\bf J^*}}$, must vanish in a Hopf bifurcation, because there is only one purely symmetric eigenvalue pair, which adds up to zero.
For detecting Hopf bifurcations we have to additionally demand that the ${\rm det~{\bf J^*}}>0$, because ${\rm tr~{\bf J^*}}=0$ is also satisfied if there is a real symmetric pair of eigenvalues, which is not characteristic of the Hopf bifurcation.
In the predator-prey model, the Hopf bifurcation is found at
\begin{equation}
\label{PPHopfBif}
s_{\rm x} = \beta d_{\rm x} + \bar{\beta} f_{\rm x} - \alpha_{\rm r} (f_{\rm y} - m_{\rm y}),
\end{equation}
where $\alpha_{\rm r}=\alpha_{\rm y}/\alpha_{\rm x}$ is the turn-over rate of the predator measured in multiples of the turnover-rate of the prey.
For systems with more than 2 variables, the testfunction for the Hopf bifurcation can be derived by a 
procedure that is described in \cite{Gross:2004p2428}.

\begin{figure*}
   \centering
   \includegraphics[width=1\textwidth]{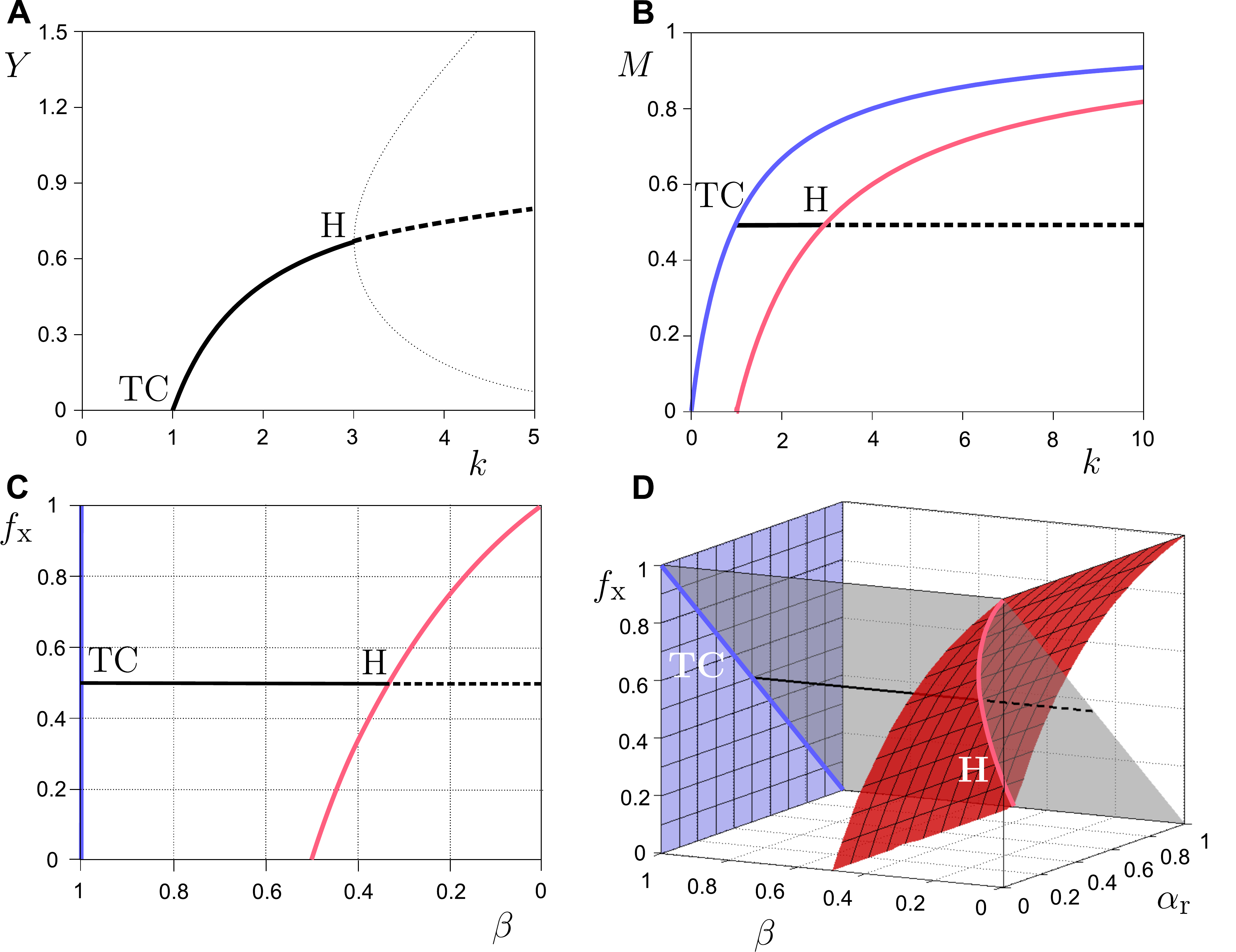}
      \caption{A. Bifurcation diagram of the Rosenzweig-MacArthur model. The predator Y can invase the system when the carrying capacity $k$ of the prey exceeds a threshold, correpsonding to a transcritical bifurcation (TC). Increasing the carrying capacity further eventually leads to destabilization in a Hopf bifurcation (H). Lines mark stable (solid) and unstable (dashed) steady states and the upper and lower turning points of a stable limit cycle (dotted). Parameters: $r=1,~a=2,~\gamma=0.5,~b=1,~m=0.5$.
B. A two-parameter bifurcation diagram of the Rosenzweig-MacArthur model as a function of the mortality of Y, $m$, and the carrying capacity of X, $k$. Stable equilibria are confined to the narrow region between the transcitical (TC, blue) and the Hopf (H, red) bifurcation points. The black lines indicate the steady states found in the section of this diagram shown in A.  
C. A two-parameter bifurcation diagram of the generalized predator-prey model as a function of the proportional mortality of X due to intra-specific competition ($\beta$) and the elasticity of the predation rate with respect to the prey ($f_{\rm x}$). Bifurcations and labels are as above. The black line plots the trajectory of the Rosenzweig MacArthur system as a specific example of the class of models.
D. A three-parameter bifurcation diagram of the generalized predator-prey model. The bifurcation points now form surfaces. The black lines indicate the steady states from A, while the grey plane indicates all steady states that can be reached in the Rosenzweig-MacArthur model if $k$ and $m$ are varied as in B.}
      \label{fig:3}
\end{figure*}

To illustrate the differences between generalized and conventional modeling we again compare the generalized model with a specific example. 
For this purpose we focus on the Rosenzweig-MacArthur model. 
In this model the prey exhibits logistic growth in absence of the predator, the predator-prey interaction is modeled by a Holling-type-II functional 
response, and the mortality of the predator is assumed to be density independent.  
This leads to
\begin{eqnarray}
&& \frac{{\rm d}}{{\rm dt}}X = rX\left(1-\frac{X}{k}\right) - \frac{aXY}{b+X}, \nonumber\\\
&& \frac{{\rm d}}{{\rm dt}}Y = \gamma \frac{aXY}{b+X} - mY, \label{eq:xdef2}
\end{eqnarray}
where $r$ is the intrinsic growth rate of X, $k$ is the carrying capacity of X, $a$ is the predation rate at saturation, $b$ is the half-saturation value of the predation rate, $\gamma$ is the biomass conversion efficiency, and $m$ is the mortality rate of Y. 

The results of a conventional bifurcation analysis are shown in Fig.~\ref{fig:3}A. If the carrying capacity $k$ is too small then the predator population 
cannot invade the system. As the carrying capacity is increased a transcritical bifurcation occurs in which a stable equilibrium appears, such that the 
predator-prey system can reside in stationarity. If the carrying capacity is increased further a supercritical Hopf bifurcation occurs, in which the equilibrium is destabilized. Subsequently, the system resides on a stable limit cycle, which emerges from the Hopf bifurcation. On this cycle pronounced predator-prey oscillations can be observed, which become larger as the carrying capacity is further increased. 

One can imagine that if an additional parameter is changed then critical values of the carrying capacity at which the bifurcations occur change
as well. This can be visualized in two-parameter bifurcation diagrams, which we have already used for the generalized model in Fig.~\ref{fig:2}B.
In such diagrams Hopf and saddle nodes bifurcation points form lines in the two-dimensional parameter space.
For the specific example of the Rosenzweig-MacArthur system, a two-parameter bifurcation diagram is shown in Fig.~\ref{fig:3}B. 
This diagram illustrates that increasing the mortality rate $m$ of the predator, shifts both the transitical bifurcation point and the Hopf bifurcation point 
to higher values of the carrying capacity.    
   
For comparing the specific example to the generalized model we compute the generalized parameters that are observed in the steady states of the specific model. Above we have already noted that logistic growth can be understood as a combination of linear reproduction and quadratic mortality, which corresponds to 
$s_{\rm x}=1$, $d_{\rm x}=2$. Furthermore, the assumptions of density independent mortality and linear dependence of the predation rate on the predator imply $m_{\rm y}=f_{\rm y}=1$.   
The elasticity $f_{\rm x}$ of the predation rate with respect to prey was derived in \cite{GrossTheorBiol} and is 
\begin{equation}
f_{\rm x} = \frac{1}{1+\chi},
\end{equation}
where $\chi = X^*/b$. Accordingly, $f_{\rm x}=1$ in the limit of vanishing prey density and $f_{\rm x}=0$ in the limit of infinite prey.
Note that in the Rosenzweig-MacArthur model the predator population tightly controls the prey population. 
Once the predator can invade, any further increase in carrying capacity only increases the stationary population of the predator, while the stationary 
population size of the prey remains invariant. 

Apart from the parameters $\beta$ and $f_{\rm x}$, shown in Fig.~\ref{fig:3}C, the only other parameter that is not fixed to a specific value is the relative turnover of the predator $\alpha_{\rm r}=\alpha_{\rm y}/\alpha_{\rm x}$. This parameter cannot affect the transcritical bifurcation, because turnover rates by construction cannot appear in testfunctions of transcritical or saddle-node bifurcations. By contrast, turnover rates in general affect Hopf bifurcations. However, in the present example the dependence of the Hopf bifurcation test function, Eq.~(\ref{PPHopfBif}) on $\alpha_r$ disappears if density independent mortality and linear dependence of the predation rate on the predator population are assumed. Therefore, the parameter has now influence on the bifurcation surfaces.

We can now map the steady states to the specific system into the generalized parameter space. 
A two-parameter bifurcation diagram of the generalized model is shown in Fig.~\ref{fig:3}C. 
In this diagram bifurcations of saddle-node type occur only on the boundary of the parameter space, where the branching parameter $\beta$ vanishes.
This parameter value indicates that none of the biomass loss of the prey occurs because of predation. 
Even without comparing to the specific example we can conclude that this bifurcation must be a transcritical bifurcation in which the predator enters the system. To illustrate this we map additionally the two-dimensional bifurcation diagram (Fig.~\ref{fig:3}B) into the generalized parameter space. This mapping is visualized in a three-dimensional bifurcation diagram shown in Fig.~\ref{fig:3}D. Such three-dimensional diagrams can be generated from analytical testfunctions using the method described in \cite{Stiefs2008}. As in the two-parameter diagrams, every point in the diagram represents a family of steady states. The parameter volume is divided by bifurcation surfaces, which separate steady states with qualitatively different local dynamics. Specifically, all steady states located between the two bifurcation surfaces are stable, whereas the steady states below the Hopf bifurcation surface are unstable.


In the present example we were able to show all relevant parameters in a single three-parameter bifurcation diagram. Let us remark that 
this is in general not possible as a larger number of parameters is often necessary to capture the dynamics of a system at the desired generality. 
Even if a generalized model contains only five parameters, the three-dimensional slice that can be visualized in a single three-parameter 
diagram is relatively small when compared to the five-dimensional space. Nevertheless, plotting three-parameter bifurcation diagrams can be very valuable 
because a three-dimensional diagram is often sufficient to locate bifurcations of higher codimension. Such bifurcations are formed at the point in parameter 
space where different bifurcation surfaces meet or intersect. The presence of such bifurcations can reveal additional insights into global properties of the dynamics. For instance in \cite{Gross:2005p2432} the presence of a certain bifurcation of higher codimension in generalized models was used to show that chaotic dynamics 
generically exist in long food chains. An extensive discussion of bifurcations of higher codimension and their dynamical implications is presented in \cite{kuznetsov2004}. For obtaining a general overview of the dynamics of larger systems containing hundreds or thousands of parameters, bifurcation diagrams are not suitable. However, these systems can be analyzed by statistical sampling techniques described in the subsequent section. 

\section{Intraguild Predation}
\label{secOmni}
As the final example we consider the effect of omnivory on a small food web. 
Omnivory is defined by an organism's ability to consume prey that inhabit multiple trophic levels.
It has been the subject of much recent interest because it is notable for its pervasiveness within well-studied ecosystems \cite{Polis:1991p1957}, as well as its relatively complex dynamics \cite{McCann:1997p2469,Kuijper:2003p2454,Tanabe:2008p2462}.
A specific case of omnivory is intraguild predation (IGP), which in its simplest incarnation appears in a three-species system containing a consumer-resource pair (as in the prior example), and an omnivore that predates upon both the consumer and resource.

Omnivory has been historically viewed as a paradoxical interaction.
Initially, the presence of omnivory was thought to be entirely destabilizing, and, as a consequence, rarely observed in nature \cite{Pimm:1978p3189}.
However, further explorations of ecological networks have reported omnivory to be a common architectural component within larger food-webs \cite{Bascompte:2005p2472,Stouffer:2010p2533}. 
Furthermore, theoretical investigations have revealed parameter regions that lead to both stabilizing and destabilizing dynamics in simple models \cite{McCann:1997p2469,Kuijper:2003p2454,Tanabe:2008p2462,Namba:2008p2458,Verdy:2010p2460}.
These theoretical arguments are limited by the fact that such models are either constrained to specific functional forms or report dynamics across parameter ranges that may not be biologically relevant.
A generalization of the entire class of simple omnivory models is poised to elucidate under which conditions stable or unstable dynamics are bound to occur, regardless of the functional relationships among or between species in the model.

We consider the generalized model
\begin{eqnarray}
&&\frac{\rm d}{\rm dt}X = S(X) -D(X) - F(X,Y) - G(X,Y,Z) \nonumber\\
&&\frac{\rm d}{\rm dt}Y = \gamma F(X,Y)-H(X,Y,Z)-M(Y) \nonumber\\
&&\frac{\rm d}{\rm dt}Z = K(X,Y,Z)-M(Z). \nonumber\\
\end{eqnarray}
In addition to the terms already present in the predator-prey model from Sec.~\ref{secPred} we included the functions $G$ and $H$, which denote the loss of the resource and consumer from predation by the omnivore and the function $K$ denoting the gain of the omnivore that arises from this predation.  
Note that we modeled the two different predatory losses of X as separate terms $G$ and $H$ because these losses can be assumed to arise independently of each other. 
By contrast the gain of the omnivore derives from predation on two different prey species and is modeled as a single term $K$ because finite handling time, saturation effects, and possibly active prey-switching behavior prevent the predator from feeding on both sources independently of each other.  

By following the procedure described in the previous sections we construct the Jacobian
\begin{equation}
{\rm \mathbf{J^*}}=
\left(
\begin{array}{ccc}
\alpha_{\rm x}(s_{\rm x} - \delta d_{\rm x} - \bar{\delta}(\beta_{\rm x} f_{\rm x} - \bar{\beta}_{\rm x} g_{\rm x})
& -\alpha_{\rm x} \bar{\delta}(\beta_{\rm x} f_{\rm y} + \bar{\beta}_{\rm x} g_{\rm y})
& -\alpha_{\rm x} \bar{\beta}_{\rm x} \bar{\delta} g_z \\
\alpha_{\rm y}(f_{\rm x} - \beta_{\rm y} h_{\rm x}
& \alpha_{\rm y}(f_{\rm y} - \beta_{\rm y} h_{\rm y} - \bar{\beta}_{\rm y} m_{\rm y})
&-\alpha_{\rm y}\beta_{\rm y}h_{\rm z}\\
\alpha_{\rm z}k_{\rm x}
& \alpha_{\rm z}k_{\rm y}
& \alpha_{\rm z}(k_{\rm z}-m_{\rm z})
\end{array}
\right)
\label{eq:genomnivJac}
\end{equation}
where the elasticities are defined as 
\begin{eqnarray}
&& s_{\rm x} := \left.\frac{\partial \log S}{\partial \log X}\right|_*, ~~ d_{\rm x} := \left.\frac{\partial \log D}{\partial \log X}\right|_*, ~~ f_{\rm x} :=  \left.\frac{\partial \log F}{\partial \log X}\right|_*,~~ g_{\rm x} :=  \left.\frac{\partial \log F}{\partial \log X}\right|_*, \nonumber\\
&& f_{\rm y} := \left. \frac{\partial \log F}{\partial \log Y}\right|_*, g_{\rm y} := \left. \frac{\partial \log G}{\partial \log Y}\right|_*,~~ h_{\rm y} := \left. \frac{\partial \log H}{\partial \log Y}\right|_*,m_{\rm y} := \left. \frac{\partial \log M}{\partial \log Y}\right|_*, \nonumber\\
&& g_{\rm z} := \left. \frac{\partial \log G}{\partial \log Z}\right|_*,~~ h_{\rm z} := \left. \frac{\partial \log H}{\partial \log Z}\right|_*,~~m_{\rm z} := \left. \frac{\partial \log M}{\partial \log Z}\right|_*, \nonumber \\
&& k_{\rm x} := \left. \frac{\partial \log K}{\partial \log X}\right|_*,~~k_{\rm y} := \left. \frac{\partial \log K}{\partial \log Y}\right|_*,~~k_{\rm z} := \left. \frac{\partial \log K}{\partial \log Z}\right|_*,
\label{eq:omnivdefs1}
\end{eqnarray}
the scale parameters are
\begin{eqnarray}
&& \alpha_{\rm x} = \frac{S^*}{X^*} = \frac{D^*}{X^*} + \frac{F^*}{X^*} + \frac{G^*}{X^*}, \nonumber \\
&& \alpha_{\rm y} = \gamma \frac{F^*}{Y^*} = \frac{H^*}{Y^*} + \frac{M^*}{Y^*},  \nonumber \\
&& \alpha_{\rm z} = \frac{K^*}{Z^*} = \frac{M^*}{Z^*},  \nonumber \\
\label{eq:omnivdefs2}
\end{eqnarray} 
and the branching parameters are
\begin{equation}
\delta = \frac{D^*}{D^*+F^*+G^*},~~ \beta_{\rm x} = \frac{F^*}{F^*+G^*},~~\beta_{\rm y} = \frac{H^*}{H^*+M^*},
\end{equation}
and $\bar{\delta} = 1 - \delta$, $\bar{\beta}_{\rm x} = 1 - \beta_{\rm x}$, and $\bar{\beta}_{\rm y} = 1 - \beta_{\rm y}$.

Let us remark that the branching parameters in the model were defined such that the parameter $\delta$ separates the predatory 
losses of the resource from the intraspecific losses. This was done to reflect our opinion that these losses are qualitatively different. An alternative procedure would have been to use three branching parameters, $\beta_d$, $\beta_f$, $\beta_g$, to denote directly the different proportions the three losses contribute to the total per-capita loss rate of X. In this case, we would have to demand $\beta_d+\beta_f+\beta_g=1$ for consistency, such that only two of the parameters could be varied independently.   

In principle the Jacobian of the omnivory model could be analyzed straight away. 
However, more insights can be gained by building more biological knowledge into the model. 
In the following we integrate this knowledge into the Jacobian derived above, by a refinement procedure that can be used to iteratively integrate 
new information into the generalized model when such information becomes available.

In the present example we want to integrate the observation that the different elasticities associated with functions describing predation by the omnivore
cannot be unrelated.   
Following the reasoning of \cite{holling} we note that the main source of nonlinearity in the predator-prey interaction is the finite handling time of captured prey. 
This handling time is principally dependent on the total amount of captured prey, which we denote by the auxilliary vairable $T$.
For simplicity, we assume that $T$ is a weighted average of prey, such that
\begin{equation}
T(X,Y) = T_{\rm x}X + T_{\rm y}Y,
\label{eq:T}
\end{equation}
where $T_{\rm x}$ and $T_{\rm y}$ are constant weights that can encode for instance different success rates for predation on the different prey species. 
In the following we denote the relative proportions that consumer and resource contribute to the diet of the omnivore as  
\begin{equation}
t_{\rm x}= T_{\rm x} \frac{X}{T(X,Y)},~~t_{\rm y}= T_{\rm y} \frac{Y}{T(X,Y)},
\label{eq:t}
\end{equation}
such that $t_x + t_y = 1$. If a species contributes a given proportion to the diet of the omnivore it is reasonable to assume that the 
same species carries an equal portion of the losses inflicted by the omnivore, such that $G(X,Y,Z) \propto t_{\rm x}K(T(X,Y),Z)$ and $H(X,Y,Z) \propto t_{\rm y}K(T(X,Y),Z)$.  

By considering these assumptions in the steady state under consideration and applying the identity Eq.~\ref{eqIdentity} we find
\begin{eqnarray}
&& g_{\rm x} = k_t t_x + t_y, \nonumber \\
&& g_{\rm y} = k_t t_y - t_y, \nonumber \\
&& h_{\rm x} = k_t t_x - t_x, \nonumber \\
&& h_{\rm y} = k_t t_y + t_x, \nonumber \\
&& k_{\rm x} = k_t t_x, \nonumber \\
&& k_{\rm y} = k_t t_y. \nonumber \\
\label{eq:Telast}
\end{eqnarray}
An exemplary derivation of one of these relations is shown in Appendix~B.  The new parameter $k_{\rm t}$ is the elasticity of the omnivore's gain with respect to the total amount of available prey, i.e. the saturation of the omnivore. This parameter can be interpreted completely analogously to the parameter $f_{\rm x}$ in the predator-prey system. 

Taking additional biological insights into account has led to relationships that can be directly substituted into the previously derived Jacobian. 
Doing so removes six parameters from the generalized model at the cost of introducing two new ones. 
The substitution makes the model less general and more specific, allowing us to extract more conclusions on a narrower range of models. 
By this procedure new insights on a given system can be integrated iteratively without reengineering the model from scratch. 
While this is clearly an academic exercise for the present three variable model, we believe that it will be valuable for future food web models possibly containing hundreds of species. 

Let us remark that iterative refinement is not contingent on the availability of a specific, i.e. non-general, equation. Instead of the specific relationship 
in Eq.~(\ref{eq:T}) we could also have used the general relationship $T(X,Y) = C_{x}(X) + C_y(Y)$, where $C_x$ and $C_y$ are general functions. Even substituting this general relationship into the model leads to a reduction of parameters of the model. Further, the functions $C_x$ and $C_y$ can be used to introduce active prey switching. This has been done for instance in the food web models proposed in \cite{Gross:2006p2337,Gross:2009p2158}.  

Using the techniques described above, the local bifurcations of the IGP model can be calculated analytically. However, because the number 
of parameters is relatively large, even three parameter diagrams reveal only a very limited insight in the dynamics of the system. 
We therefore use an alternative approach and explore the parameter space by a numerical sampling procedure.
Because all parameters in the model have clear interpretation, we can assign a range of realistic values to each of the parameters.
We generate an ensemble of parameter sets by randomly assigning each parameter a value drawn from the respective range.
The stability of the steady state corresponding to a sample parameter set is then determined by numerical computation of the eigenvalues of the 
corresponding Jacobian. Because of the high numerical efficiency of eigenvalue computation, ensembles of millions or billions of
sample parameter sets can be evaluated in reasonable computational time. 
Based on such large ensembles, a sound statistical analysis of models containing hundreds or thousands of parameters is feasible. 
An example of such an analysis in a 50 species model was presented in \cite{Gross:2009p2158}.

\begin{table}
\centering
\begin{tabular}{|c|c|c|}
\hline parameter & value or range \\ 
\hline $\alpha_{\rm x}$& 1 \\ 
$\alpha_{\rm y}$ & $r$ \\ 
$\alpha_{\rm z}$ & $r^2$ \\ 
$r$& 0 to 1 \\ 
$s_{\rm x}$& 1 \\ 
$d_{\rm x}$& 2 \\ 
$m_{\rm y}$& 1 \\ 
$m_{\rm z}$& 1 \\ 
$f_{\rm y}$& 1 \\ 
$f_{\rm x}$& 0 to 1 \\ 
$k_{\rm t}$& 0 to 1 \\
$k_{\rm z}$& 1 \\ 
$\delta$& 0 to 1 \\
$\beta_{\rm x}$& 0 to 1 \\
$\beta_{\rm y}$& 0 to 1 \\ 
\hline 
\end{tabular}
\caption{Values and ranges of the parameter sampling assumed to compute Fig.~\ref{fig:5}. The timescales, $\alpha_{\rm x}$, $\alpha_{\rm y}$ and $\alpha_{\rm z}$, are assumed to scale allometrically. }
\label{Tab1}
\end{table}

To assess the dependence of the stability of the IGP model on the parameters, we generated $10^8$ random parameter sets, in which the parameter values were drawn independently from uniform distributions (see~Tab.~\ref{Tab1}).
Subsequently each parameter set was assigned a stability value of 1 if it is found to correspond to a stable steady state and 0 if found to correspond to an unstable steady state. 
The dependence of stability on individual parameter values was then quantified by computing the correlation coefficient between a given parameter and the stability value over the whole ensemble.
Strong positive correlations indicate that large values of the respective parameter promote stability, while strong negative correlations indicate that large values of the parameter reduces stability.

\begin{figure*}
   \centering
   \includegraphics[width=1\textwidth]{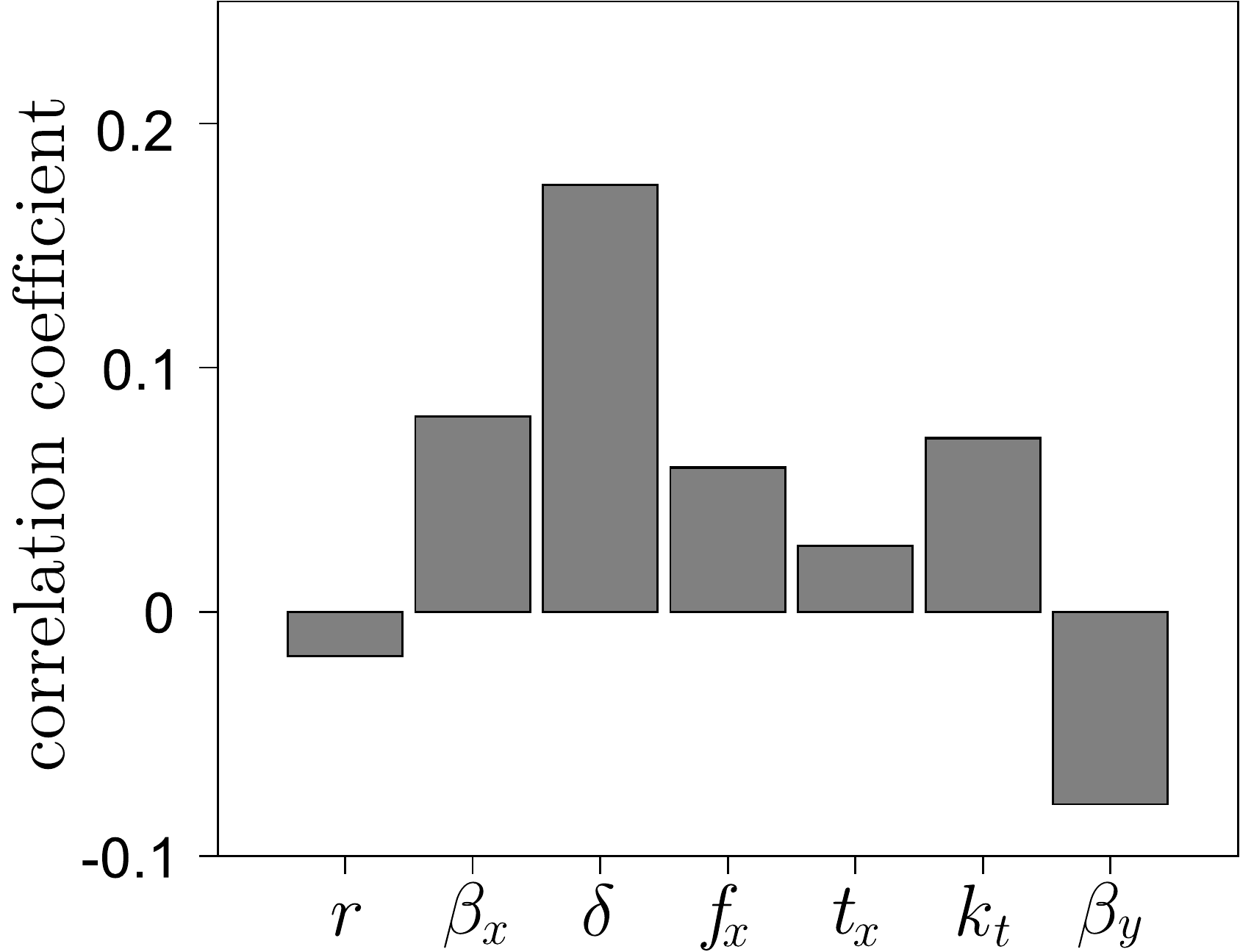}
      \caption{The dependence of the stability of the generalized IGP model on the parameters: $r, \beta{\rm x},~\delta,~f_{\rm x},~t_{\rm x},~k_{\rm t},~\beta_{\rm y},~\mbox{and}~m_{\rm z}$. Error bars are too small to be shown. Strong positive correlations indicate that large values of the specific parameter promote stability, while strong negative correlations indicate that large values of the specific parameter hinder stabilitry. 
}
\label{fig:5}
\end{figure*}

The results of the numerical analysis (Fig. \ref{fig:5}) show that the proportional loss of X due to intra-specific competition, $\delta$, and to a lesser extent, the proportional predation mortality of X due to Y, $\beta_{\rm x}$, is strongly correlated with stability. This suggests that strong competitive effects of the resource, as well as a weak omnivory interaction between X and Z increases the likelihood of stability within the IGP system. Conversely, $\beta_{\rm y}$, the proportional mortality of Y due to extrinsic factors, shows a strong negative correlation with stability. 
This suggests that a strong interaction between Y and Z (low $\beta_{\rm y}$) facilitates stability.

We remark that the precise results of the sampling analysis used here, are not indepent of the specific ranges and distributions that are used for generating 
the ensemble. Although the error bars of the statistical analysis rapidly become very small, minor differences between correlation coefficient should not be overinterpretaed. Nevertheless, the stability correlation analysis is a powerful tool that can very quickly convey an impression of the stabilizing and destabilizing factors in large networks. Ideally this analysis should be followed up by more refined statistical exploration of the ensemble. More detailed insights in the behavior of the system can be gained for instance by plotting histograms of the proportion of stable states that are found if one parameter is set to a specific value, while all others are varied randomly. Such histograms have for instance been used in \cite{RalfSteuer08082006,Steuer2007,Gross:2009p2158,Zumsande2010a}.
Because these more detailed analyses clearly exceed the scope of the present paper, we postpone further analysis of the IGP model to a separate publication.  

\section{Conclusions}
In the present paper we have illustrated the fundamental ideas or procedures of generalized modeling and extended the approach of generalized modeling.
Generalized models can reveal conditions for the stability of steady states in large classes of systems, identify the bifurcations in which stability is lost, and provide some insights into the global dynamics of the system.
They can be seen as an intermediate approach that has many advantages of conventional equation-based models, while coming close to the efficiency of random matrix models. 
This efficiency, both in terms of manual labor and CPU time, highlights generalized modeling as a promising approach for detailed analysis of large ecological systems. 
Although we have restricted the presentation to models with up to three variables, these simple examples contain already all of the complexities that are encountered in larger systems, such as the 50-species model studied in \cite{Gross:2009p2158}.

The presentation of generalized modeling in the present paper differed significantly from previous publications. 
The differences arise in part from the stronger focus on fundamental concepts and modeling strategies and in part from a newly proposed shortcut 
that facilitates the formulation of generalized models 

Throughout this paper we have contrasted several generalized models with conventional counterparts.
We emphasize that this was done purely for illustration of the results of generalized modeling. Generalized modeling should by no means regarded as an alternative modeling approach replacing conventional models. Note that generalized modeling is mainly useful in systems for which little information is available, whereas in well-known systems many more insights may be extractable by conventional models. We point out that the iterative refinement procedure proposed here, allows a researcher to start out with a generalized model and then successively integrate new information as it becomes available until eventually a conventional model is obtained. Generalized modeling should therefore be considered as a high-throughput screening tool for potential models, that is used ideally before conventional modeling of a given system is attempted.


\section{Appendix A.}
\label{secAppendixA}
The formulation of elasticity parameters is contingent on the relationship
\begin{equation}
\label{eqtoprove}
\left. \frac{\partial F}{\partial X} \right|_* = \frac{F^*}{X^*} \left.\frac{\partial \log F}{\partial \log X}\right|_*
\end{equation}
where $F$ represents some function of $X$. For proving this relationship we consider the right hand side and multiply it by $1=\partial F /\partial F$
\begin{equation}
\frac{F^*}{X^*} \frac{\partial \log F}{\partial F} \left.\frac{\partial F}{\partial \log X}\right|_*,
\end{equation}
where $(\partial \log F / \partial F)_*$ simplifies to $1/F^*$. This results in
\begin{equation}
\left. \frac{\partial F}{\partial X} \right|_* = \frac{1}{X^*} \left. \frac{\partial F}{\partial \log X} \right|_*.
\end{equation}
In  the previous steps we replaced $\log F$ in the numerator of the derivative. To replace $\log X$ in the numerator we proceed analogously   
\begin{equation}
\label{eqlong}
\left. \frac{\partial F}{\partial X} \right|_* = \frac{1}{X^*} \frac{\partial X}{\partial \log X} \left. \frac{\partial F}{\partial X} \right|_*,
\end{equation}
We now consider the second factor of the right-hand side. To evaluate the partial derivative we define $X = {\rm e}^u$ and write
\begin{equation}
\left. \frac{\partial X}{\partial \log X}\right|_* = \left. \frac{\partial {\rm e}^u}{\partial \log {\rm e}^u} \right|_* =  \left. \frac{\partial {\rm e}^u}{\partial u}\right|_* = \left. {\rm e}^u\right|_*=X^*.
\end{equation}
substituting back into Eq.~(\ref{eqlong}) we obtain
\begin{equation}
\left. \frac{\partial F}{\partial X^*} \right|_* =\frac{1}{X^*} X^* \left. \frac{\partial F}{\partial X} \right|_* = \left. \frac{\partial F}{\partial X} \right|_*.
\end{equation}
which proves Eq.~(\ref{eqtoprove}).

\section{Appendix B.}
\label{secAppendixB}
Starting from 
\begin{equation}
G(T(X,Y),Z) = \frac{\epsilon_g T_X X}{T(X,Y)} K(T(X,Y),Z)
\end{equation}
the elasticity with respect to $X$ can be subsequently be formulated as
\begin{eqnarray}
&g_{\rm x}&=\frac{\partial \log G}{\partial \log X}\nonumber\\\
&&=\frac{X^\ast}{G^\ast} \frac{\partial}{\partial X} \frac{\epsilon_g T_x X}{T(X,Y)} K(T(X,Y),Z) \nonumber\\\
&&=\frac{\epsilon_g X^\ast T^\ast}{\epsilon_g T_x X^\ast K^\ast}\left( \frac{T_x K^\ast}{T^\ast} + \frac{T_x X^\ast}{T^\ast}\frac{\partial K}{\partial X} + T_x X^\ast K^\ast \frac{\partial}{\partial X}\frac{1}{T(X,Y)}\right) \nonumber\\\
&&=\frac{T^\ast}{T_x K^\ast}\left( \frac{T_x K^\ast}{T^\ast} + \frac{T_x K^\ast}{T^\ast} \frac{X^\ast}{K^\ast}\frac{\partial K}{\partial X} +\frac{T_x K^\ast}{T^\ast}T^\ast X^\ast \left(-\frac{1}{(T^\ast)^2}\frac{\partial T}{\partial X}\right)\right) \nonumber\\\
&&= 1 + \frac{T^\ast}{K^\ast}\frac{\partial K}{\partial T}\frac{X^\ast}{T^\ast}\frac{\partial T}{\partial X}-\frac{X^\ast}{T^\ast} \frac{\partial T}{\partial X} \nonumber\\
&&= 1 + k_{\rm t} t_{\rm x} - t_{\rm x} \nonumber\\\
&& = k_{\rm t} t_{\rm x} + t_{\rm y}. \nonumber\\\
\end{eqnarray}



%

\end{document}